**Switching of perpendicular magnetization by spin-orbit torques in the absence of external magnetic fields**

Guoqiang Yu[1]†*, Pramey Upadhyaya[1]†, Yabin Fan[1], Juan G Alzate[1], Wanjun Jiang[1], Kin L. Wong[1], So Takei[2], Scott A Bender[2], Murong Lang[1], Jianshi Tang[1], Yaroslav Tserkovnyak[2], Pedram Khalili Amiri[1*] and Kang L. Wang[1*]

[1]Department of Electrical Engineering, University of California, Los Angeles, California 90095, United States

[2]Department of Physics and Astronomy, University of California, Los Angeles, California 90095, United States

†These authors contributed equally to this work

email address: guoqianyu@ucla.edu, pedramk@ucla.edu, wang@seas.ucla.edu

**Magnetization switching by current-induced spin-orbit torques (SOTs) is of great interest due to its potential applications for ultralow-power memory and logic devices. In order to be of technological interest, SOT effects need to switch ferromagnets with a perpendicular (out-of-plane) magnetization. Currently, however, this typically requires the presence of an in-plane external magnetic field, which is a major obstacle for practical applications. Here we report for the first time on SOT-induced switching of out-of-plane magnetized Ta/$Co_{20}Fe_{60}B_{20}$/$TaO_x$ structures without the need for any external magnetic fields, driven by in-plane currents. This is achieved by introducing a lateral structural asymmetry into our devices during fabrication. The results show that a new field-like SOT is induced by in-plane currents in such asymmetric structures. The direction of the current-induced effective field corresponding to this new field-like SOT is out-of-plane, which facilitates switching of perpendicular magnets. This work thus provides a pathway towards bias-field-free SOT devices.**

Breaking of structural symmetries of nanomagnetic systems is of great interest for the development of ultralow-power spintronic devices. The structural asymmetry in various magnetic heterostructures has been engineered to reveal novel fundamental interactions between electric currents and magnetization, resulting in spin-orbit-torques (SOTs) on the magnetization[1-6], which are both fundamentally important and technologically promising for device applications. Such SOTs have been used to realize current-induced magnetization switching[2-4, 7] and domain-wall motion[8-10] in recent experiments. Typical heterostructures exhibiting SOTs consist of a ferromagnet (F) with a heavy nonmagnetic metal (NM) having strong spin-orbit coupling on one side, and an insulator (I) on the other side (referred to as NM/F/I structures, shown schematically in Fig. 1a, which break mirror symmetry in the growth direction). In terms of device applications, the use of SOTs in NM/F/I structures allows for a significantly lower write current compared to regular spin-transfer-torque (STT) devices[4]. It can greatly improve energy efficiency and scalability[1-5, 11] for new SOT-based devices such as magnetic random access memory (SOT-MRAM), going beyond state-of-the-art STT-MRAM.

For practical applications, a critical requirement to achieve high-density SOT memory is the ability to perform SOT-induced switching *without the use of external magnetic fields*, in particular for perpendicularly-magnetized ferromagnets, which show better scalability and thermal stability as compared to the in-plane case[12]. However, there are currently no practical solutions that meet this requirement. In NM/F/I heterostructures studied so far, the form of the resultant current-induced SOT alone does not allow for deterministic switching of a perpendicular ferromagnet, requiring application of an additional external in-plane magnetic field to switch the perpendicular magnetization[2-4]. (This is a very general feature of SOT devices, which can be explained by symmetry-based arguments, as discussed below). In such experiments, the external field allows for each current direction to favor a particular orientation for the out-of-plane component of magnetization, thereby resulting in deterministic perpendicular switching. However, this external field is undesirable from a practical point of view. For device applications, it also reduces the thermal stability of the perpendicular magnet by lowering the zero-current energy barrier between the stable perpendicular states, resulting in a shorter retention time if used for memory.

This work provides a solution to eliminate the use of external magnetic fields, bringing SOT-based spintronic devices such as SOT-MRAM closer to practical application. We present a new NM/F/I structure, which provides a novel spin-orbit torque, resulting in zero-field current-induced switching of perpendicular magnetization. Our device consists of a stack of $Ta/Co_{20}Fe_{60}B_{20}/TaO_x$ layers, but also has a *structural mirror asymmetry along the in-plane direction*. The lateral structural asymmetry, in effect, replaces the role of the external in-plane magnetic field. We present experimental results on current-induced SOT switching of perpendicular magnetization without applied magnetic fields. We also present a symmetry-based analysis of SOT interactions and show that this type of bias-field-free switching originates from the lateral symmetry-breaking of the device, which gives rise to a new field-like torque upon application of an in-plane current. This new type of torque brings SOT-devices substantially closer to practical applications.

## Symmetry-based analysis of current-induced spin-orbit torques

The current-induced SOT terms, which are physically allowed for a particular device structure, can be determined based on its symmetry properties. Hence, symmetry arguments provide a powerful tool in designing magnetic material and device structures to realize particular switching characteristics. Figure 1 schematically illustrates how lateral symmetry-breaking in the device can give rise to current-induced switching of perpendicular magnetization. The coordinate system is chosen such that the $z$-axis is fixed along the growth direction, and the current is applied along the $x$-axis. The figure depicts the following scenarios:

(i) ***Mirror symmetry-breaking only along z-axis***: Figure 1a depicts the case where mirror symmetry is broken only along the $z$-axis (*i.e.* preserving mirror symmetries along the $x$- and $y$-axes similar to previous works[2-4, 7]), in the absence of external magnetic fields. In this case, the symmetry breaking results in current-induced SOTs, which, to quadratic order in $\boldsymbol{m}$, consist of a field-like (FL) term $\boldsymbol{T_c^{\mathrm{FL}}} = \gamma H_y^{\mathrm{FL}} \boldsymbol{m} \times \boldsymbol{y}$ and a damping-like (DL) term $\boldsymbol{T_c^{\mathrm{DL}}} = \gamma H_y^{\mathrm{DL}} \boldsymbol{m} \times \boldsymbol{m} \times \boldsymbol{y}$[13]. Here, $\boldsymbol{m}$ denotes a unit vector along the magnetization direction. Equivalently, these torques can be expressed in terms of effective magnetic fields, namely $\boldsymbol{H_y^{\mathrm{FL}}} = H_y^{\mathrm{FL}} \boldsymbol{y}$ and $\boldsymbol{H_y^{\mathrm{DL}}} = H_y^{\mathrm{DL}} \boldsymbol{m} \times \boldsymbol{y}$ (depicted in Fig. 1), with $H_y^{\mathrm{FL}}$ and $H_y^{\mathrm{DL}}$ representing the current-dependent proportionality constants for each term. Within simple microscopic models[14, 15], such SOTs can be derived from the interfacial Rashba interaction[16] and/or the spin Hall effect in the normal metal[17-20]. In this case, the absence of deterministic perpendicular magnetization switching can be understood by performing a mirror reflection with respect to the $xz$ plane on the state with perpendicular magnetization component $M_z > 0$, as illustrated in Fig. 1a. Under such a transformation, the direction of the current density $J$ is unaltered. However, magnetization (being a pseudo-vector) reverses the direction of its components that are parallel to the $xz$ plane, hence resulting in an equilibrium state with $M_z < 0$ in the mirror state. As a result, if a particular direction of current allows an equilibrium magnetization state with a positive $z$ component, *i.e.*, $M_z > 0$, the same direction of current should also favor a state with $M_z < 0$. In consequence, a given current direction does not favor a unique perpendicular magnetization orientation, and hence no deterministic switching is obtained[21]. This argument suggests that in order to achieve current-induced switching of perpendicular magnetization, the mirror symmetry with respect to the $xz$ plane also has to be broken. This has been achieved in previous works using an external magnetic field $H_{ap}$ [3, 7] along the current direction, as depicted in Fig. 1b. The mirror transformation in this case also reverses the external magnetic field direction. Thus, by fixing the direction of the external field along the current direction, the symmetry between magnetic states with opposite $z$-components of magnetization is broken, allowing for a unique magnetic state. (For the case shown, a positive/negative external field favors the state with positive/negative $M_z$). Formally, such a scenario has also been explained by solving for the equilibrium magnetization orientation in the presence of SOTs within a single-domain model[3].

**(ii)** ***Mirror symmetry-breaking along both z- and y-axes***: When the mirror symmetry along the $y$-axis is also broken, a particular direction of current can uniquely determine the $z$-component of magnetization. This is illustrated in Fig. 1c, where the structural asymmetry along the $y$-axis consists of a varying thickness (*i.e.* wedge shape) of the insulating layer along this axis. The mirror transformation in this case reverses both the direction of $M_z$ and the direction of $J$ (with respect to the wedge), hence associating each current direction with a unique orientation of $M_z$. In this sense, breaking structural inversion symmetry along the lateral direction can replace the role of the external bias field. This fact is also reflected in the form of the allowed current-induced SOT terms. Using a symmetry-based phenomenology[22], the current-induced SOT terms (up to quadratic order in $\boldsymbol{m}$) arising due to mirror asymmetry along both $y$- and $z$-axes can be written as (see Supplementary Materials for details)

$$\boldsymbol{T}_{SOT} = \gamma H_y^{\mathrm{FL}} \boldsymbol{m} \times \boldsymbol{y} + \gamma H_y^{\mathrm{DL}} \boldsymbol{m} \times \boldsymbol{m} \times \boldsymbol{y} + \gamma H_z^{\mathrm{FL}} \boldsymbol{m} \times \boldsymbol{z} + \gamma H_z^{\mathrm{DL}} \boldsymbol{m} \times \boldsymbol{m} \times \boldsymbol{z} \quad (1)$$

Here, in the last two terms $H_z^{\mathrm{FL}}$ and $H_z^{\mathrm{DL}}$ parameterize the strengths of the current-induced effective fields arising from the additional inversion asymmetry, respectively representing the new FL and DL SOT terms. The new FL term gives rise to a current-induced effective field ($H_z^{\mathrm{FL}}$) along the $z$-axis (depicted in Fig. 1c) and can thus facilitate current-induced deterministic switching of perpendicular magnetization in the absence of an external magnetic field[23]. In the following, we provide an experimental demonstration of this new SOT-induced perpendicular effective field and switching.

## <u>Experiment</u>

Experiments were carried out on sputter-deposited Ta(5.0 nm)/$Co_{20}Fe_{60}B_{20}$(1.0 nm)/$TaO_x$(wedge) films. The top oxide layer was formed by first depositing a Ta film, the thickness of which was varied across the wafer, as shown in Fig. 2a. The $TaO_x$ was then formed by exposing the sample to a radio-frequency $O_2$/Ar plasma, to create a $Co_{20}Fe_{60}B_{20}/TaO_x$ interface. Due to the variation of the top Ta layer thickness, the thickness of the resulting oxide as well as the oxygen content at the $Co_{20}Fe_{60}B_{20}/TaO_x$ interface change continuously across the wafer. The film was then patterned into an array of Hall bars, the structure of which is shown in Figs. 2a and 2b. The transverse direction of the Hall bars (*i.e.*, the *y*-axis) is along the $TaO_x$ wedge, as shown in Fig. 2a, breaking the mirror symmetry with respect to the *xz* plane. Thus, application of a current along the Hall bars is expected to produce an out-of-plane effective magnetic field, based on the symmetry arguments discussed above.

The devices were characterized using extraordinary Hall effect (EHE) measurements, as shown in Fig. 2b. The effective perpendicular anisotropy field ($H_k$) of the $Co_{20}Fe_{60}B_{20}$ layer (in the absence of current-induced SOT) was determined using EHE measurements as a function of the applied in-plane magnetic field. Fig. 2c shows the measured $H_k$ as a function of position along the $TaO_x$ gradient direction. The curve shows a non-monotonic dependence of the perpendicular magnetic anisotropy (PMA) on position, indicating an increase of $H_k$ on the thinner side ($dH_k/dy > 0$) and decrease of $H_k$ on the thicker side ($dH_k/dy < 0$) of the wedge. The perpendicular magnetization, measured as a function of perpendicular magnetic field, is shown in Fig. 2d. For devices located on the central region of the wedge with the largest $H_k$, the EHE perpendicular loops are square-shaped and show a large coercivity. As expected, the loops become less square-shaped and eventually turn into hard-axis-like loops on both sides of the wedge where $H_k$ is smaller. The observed PMA is due to the interfacial magnetic anisotropy between the $Co_{20}Fe_{60}B_{20}$ film and its adjacent $TaO_x$ and Ta layers, similar to recent reports in other material systems[24-26]. The anisotropy associated with the $TaO_x$ interface is in turn affected by the appearance of Fe-O and Co-O bonds at the interface[25, 26], exhibiting a non-monotonic dependence on the oxygen content[27, 28]. As a result, the change of PMA across the wedge reflects the gradient of oxygen concentration at the $Co_{20}Fe_{60}B_{20}/TaO_x$ interface across the wafer.

We next performed EHE measurements on the Hall bar devices for a set of different direct currents applied along the $x$-axis. Figs. 3a-c show the measured EHE signals for one device (device A, $t_{Ta}$ = 1.65 nm prior to oxidation) in the $dH_k/dy > 0$ region. As expected, small currents have almost no influence on the switching behavior, as shown in Fig. 3a. At larger currents, however, the centers of the hysteresis loops are gradually shifted to the left for currents of a positive polarity, which indicates the presence of a perpendicular effective field, $\boldsymbol{H}_z^{\mathrm{FL}} = H_z^{\mathrm{FL}}\boldsymbol{z}$ induced by the current. The value of $H_z^{\mathrm{FL}}$ can be extracted from the average of the positive ($H_S^+$) and negative ($H_S^-$) switching fields, *i.e.* $H_z^{FL} = -(H_S^+ + H_S^-)/2$. For currents in the opposite direction, the hysteresis loops are shifted to the right. At current values of $I$ = ±10 mA, the separation between the two loops for this device (Fig. 3c) is $H_z^{FL}$(10 mA) − $H_z^{FL}$(−10 mA) ≈ 22 Oe. It is interesting to compare this to the measurements shown in Figs. 3d-f, which show the EHE signal for a different device (device B, $t_{Ta}$ = 1.94 nm prior to oxidation) in the $dH_k/dy < 0$ region (*i.e.* on the opposite side of the anisotropy peak in Fig. 2c). In this case, the sign of the current-induced field is opposite to device A for the same direction of current flow. The separation between the two loops at $I$ = ±10 mA is $H_z^{FL}$(10 mA) − $H_z^{FL}$(−10 mA) ≈ −66 Oe for device B. Thus, by comparing the $H_z^{\mathrm{FL}}$ of these two devices, it is evident that $H_z^{\mathrm{FL}}(I > 0)$ − $H_z^{\mathrm{FL}}(I < 0) > 0$ [*i.e.* $dH_z^{\mathrm{FL}}/dI > 0$] in the d$H_k$/dy > 0 region, while $H_z^{\mathrm{FL}}(I > 0) - H_z^{\mathrm{FL}}(I < 0) < 0$ [*i.e.* $dH_z^{\mathrm{FL}}/dI < 0$] in the d$H_k$/dy < 0 region.

To quantify the $H_z^{\mathrm{FL}}$ induced by current, the values of $H_S^+$ and $H_S^-$ for the two devices (obtained from EHE loops in Fig. 3), are summarized in Figs. 4a and b, for different applied currents. The current-induced perpendicular field can then be obtained by fitting the current dependence of $H_z^{FL} = -(H_S^+ + H_S^-)/2$. Here, Fig. 4a corresponds to device A (d$H_k$/dy > 0), and Fig. 4b corresponds to device B (d$H_k$/dy < 0). For both cases, the resultant $H_z^{\mathrm{FL}}$ can be fitted well to a linear curve, and hence can be expressed as $H_z^{\mathrm{FL}} = \beta J$, where $J$ is the applied current density. The values of $\beta$, extracted in a similar fashion for all devices measured along the wedge, are shown in Fig. 4c, with the largest absolute value of $\beta$ reaching ~ 56 Oe per $10^{11}$Am$^{-2}$. The plot also shows d$H_k$/dy as a function of position along the wedge for comparison. It can be seen that both the sign and magnitude of $\beta$, and hence $H_z^{\mathrm{FL}}$, correspond well with d$H_k$/dy. Thus, in addition to the sign and magnitude of the applied current, the current-induced $H_z^{\mathrm{FL}}$ also depends on the sign and magnitude of d$H_k$/dy. This, in turn, establishes the correlation of $H_z^{FL}$ with the symmetry-breaking along the *y*-axis, confirming the central hypothesis of this work.

## Discussion

The symmetry argument presented above does not provide any details on the microscopic mechanisms behind the creation of the new perpendicular effective field. Hence, in addition to spin-orbit effects, in principle current-induced magnetic (Oersted) fields may also contribute to the observed perpendicular switching.

However, this possibility can be ruled out in our devices, as will be explained next. Due to the nonuniform oxidation of the $TaO_x$ layer, the structural asymmetry may cause a nonuniform current density along the width of our devices. Hence, there will be a larger current density on one side (less oxidized part) of the Hall bar, producing a net perpendicular magnetic field within the Hall bar area. However, for a particular current direction, this kind of Oersted field would be expected to point in the same direction for *all* Hall bars, since they all have an identical direction of the $TaO_x$ thickness gradient. Hence, based on the asymmetric sample structure shown in Fig. 1a, the Oersted field should result in a negative sign of $\beta$ for all devices, which is not the case observed in our experiments. This indicates that Oersted fields are not the origin of the observed $H_z^{FL}$ in our samples. In addition, we estimated the Oersted field induced by current in our structures (see Supplementary Materials). The estimation indicates that, the value of the Oersted field induced by current is ~16 times smaller than the largest current-induced perpendicular shift (induced by $H_z^{\mathrm{FL}}$) observed in our experiments. Hence, the role of Oersted fields can be excluded, and thus we attribute the origin of the observed perpendicular loop shifts to SOT.

In addition, we also ruled out the possibility of a multi-domain process playing a role in the observed FL shifts in the hysteresis loops. This was done by measuring $H_z^{FL}$ *via* independent second-harmonic measurements[5, 13, 29] in the presence of large external magnetic fields. The applied field was larger than the saturation field of our samples, ensuring single-domain behavior during the experiment. The results show an excellent agreement between the current-induced $H_z^{FL}$ extracted from second-harmonic measurements and the hysteresis loop shifts (see Supplementary Materials for details).

We further compared the magnitude of the new perpendicular effective field $H_z^{\mathrm{FL}}$, with the regular effective field $H_y^{\mathrm{FL}}$, which results from the of inversion asymmetry along the *z*-axis[1, 5, 9, 30]. The largest magnitude of $H_z^{\mathrm{FL}}$ in our samples is 4.7 Oe/mA, ≈ 56 Oe per $10^{11}$ $\mathrm{Am}^{-2}$. The $H_y^{\mathrm{FL}}$ as measured by the second-harmonic method, on the other hand, is 170 Oe per $10^{11}$ $\mathrm{Am}^{-2}$ for the same device (see Supplementary Materials for details). Thus, the perpendicular effective field $H_z^{\mathrm{FL}}$ is of sizable strength when compared to the regular (in-plane) field-like term induced by breaking of inversion symmetry along the *z*-axis. However, the presence of this additional $H_z^{\mathrm{FL}}$ provides significant advantages for device applications, as it enables bias-field-free switching of out-of-plane magnetic layers.

Microscopically, the new field-like torque appears to stem from the lateral oxidation gradient at the $Co_{20}Fe_{60}B_{20}/TaO_x$ interface, which can induce Rashba-like spin-orbit coupling with the effective electric field direction pointing along the wedging direction *y*. Namely, a microscopic electron Hamiltonian of the form $\mathcal{H} \sim \boldsymbol{\sigma} \cdot (\boldsymbol{y} \times \boldsymbol{p})$[16], at the $Co_{20}Fe_{60}B_{20}/TaO_x$ interface, could in principle account for a field-like torque of the form $\boldsymbol{T^{FL}} \sim \boldsymbol{m} \times \boldsymbol{H_z^{FL}} \sim \boldsymbol{m} \times (\boldsymbol{y} \times \boldsymbol{J})$. Here $\boldsymbol{\sigma}$ and $\boldsymbol{p}$ stand for the Pauli matrices and the electron's momentum operator, respectively. As shown in Fig. 1d, an electric field along the wedging direction could, in turn, originate from redistribution of charges near the interface depending on the oxygen content, which is also responsible for the non-monotonic dependence of $H_k$ on position[28]. However, a more detailed understanding of these electric fields and their possible contribution to $H_z^{\mathrm{FL}}$ is needed and requires first principles calculations.

**Current-induced switching in the absence of external fields**

Finally, we demonstrate that the perpendicular effective field $H_z^{\mathrm{FL}}$, induced by an in-plane current along the $x$-axis, can be used to deterministically switch the magnetization in an out-of-plane magnetized film, without assistance of external magnetic fields. This is shown in Figs. 5a and b, for two representative devices on different sides of the $H_k$ peak in Fig. 2c. We were able to reversibly switch the perpendicular magnetization by currents of ~6 mA (corresponding to a current density of $5.0 \times 10^6$ A/cm$^2$) for device C ($t_{\mathrm{Ta}}$ = 1.67 nm before oxidation) in Fig. 5a, and by current of ~3 mA (corresponding to a current density of $2.5 \times 10^6$ A/cm$^2$) for device B in Fig. 5b. For the first device, which is in the d$H_k$/dy > 0 region, positive currents favor a positive magnetization (resulting in a negative Hall resistance $R_{Hall}$). For the latter device, which is in the d$H_k$/dy < 0 region, positive currents favor a negative magnetization. The favored direction of magnetization for a particular current direction is dependent on the sign of $\beta$, and hence depends on the location along the wedge, as expected. Thus, currents of opposite polarities can be used to switch the perpendicular magnetization in opposite directions, and the favored direction of magnetization for each current is determined by the sign of the lateral device asymmetry, as quantified by the sign of d$H_k$/dy. We obtained similar results for several other devices measured at different points along the wedge. Due to the correlation between d$H_k$/dy and the strength of $H_z^{\mathrm{FL}}$ (see Fig. 4c), we expect that the current density required for switching can be further reduced by increasing the value of $dH_k/dy$, *i.e.* by creating a larger structural asymmetry in the device.

For device applications, the perpendicular $H_z^{\mathrm{FL}}$ induced by currents in this work can be used in three-terminal structures, where the perpendicular ferromagnetic free layer is part of a magnetic tunnel junction, allowing for readout of its state *via* the tunneling magnetoresistance (TMR) effect. No external in-plane magnetic fields would be needed to operate the device. It should also be noted that the nonuniform oxidation method used to create the lateral asymmetry in this work is not the only approach that could be used for this purpose. For integration into large device arrays (*e.g.* for memory chips), which requires uniformity across the wafer, a more localized method of generating the lateral asymmetry may be more appropriate. We expect that this work will motivate research to develop such asymmetric device structures, which exhibit and utilize perpendicular effective field. Examples of alternative approaches to create these torques could include the application of a lateral voltage to the device, using effects such as voltage-controlled magnetic anisotropy, or to use a nonuniform strain built into the device during the fabrication process, all of which could be used to break the lateral inversion symmetry. By providing large perpendicular effective field, such structures can result in new ultralow-power and highly scalable SOT-based spintronic memory and logic circuits.

**Methods**

The stack structure of $Ta/CoFeB/TaO_x$ was fabricated from Ta(5 nm)/$Co_{20}Fe_{60}B_{20}$(1 nm)/Ta(wedge) sputtered films. The metal layers were deposited on a thermally oxidized wafer (on an area of 10 mm × 50 mm) by d.c. magnetron sputtering at room temperature, in an AJA international physical vapor deposition system. The deposition rates were 0.06 nm/s for Ta and 0.03 nm/s for $Co_{20}Fe_{60}B_{20}$ at an argon pressure of 2 mTorr and 3 mTorr, respectively. The top Ta was grown in a wedge shape, giving a continuous gradient of thickness along the length of the sample. The thickness of the top Ta was varied from 0.81 to 2.13 nm. The $TaO_x$ layer was then formed by exposing the sample to a radio-frequency $O_2$/Ar plasma for 100 s. The top wedged Ta layer was thus oxidized, resulting in a change of oxidation level at the $CoFeB/TaO_x$ interface along the wedge direction. The films were then annealed at 200 ℃ for 30 min to enhance their perpendicular magnetic anisotropy (PMA). The sample was subsequently patterned into an array of Hall bar devices (seven in the width direction, with constant thickness of the top Ta layer, and thirty-five in the length direction of the sample, varying its thickness) by standard photolithography and dry etching techniques. The size of the Hall bars was fixed at 20 $\mu$m × 130 $\mu$m. The Hall bar lengths were oriented along the width direction of the sample, resulting in a varying top Ta thickness (hence oxidation) along the width of the Hall bars (*i.e.* *y*-axis). The spacing between two adjacent Hall bar devices in the length direction was 1.1 mm. A Keithley 6221 current source and a Keithley 2182A nano-voltmeter were used in the extraordinary Hall voltage measurement. The external magnetic field was generated by a Helmholtz coil driven by a Kepco power supply. All measurements were carried out at room temperature.

**Acknowledgements**

This work was partially supported by the DARPA program on Nonvolatile Logic (NVL), and in part by the NSF Nanosystems Engineering Research Center for Translational Applications of Nanoscale Multiferroic Systems (TANMS). This work was also supported in part by the FAME Center, one of six centers of STARnet, a Semiconductor Research Corporation (SRC) program sponsored by MARCO and DARPA. P.U. and J.G.A. acknowledge partial support from a Qualcomm Innovation Fellowship.

**Author contributions**

G.Q.Y. and P.U. jointly conceived the idea with contributions from Y.T., P.K.A. and K.L.W.. G.Q.Y. designed the experiments to test the idea, fabricated and measured the devices with contributions from Y.F., J.G.A., W.J., K.W., M.L. and J.T.. P.U. performed the theoretical analysis and modeling with help from S.T., S.A.B. and Y.T.. G.Q.Y., P.U., P.K.A. and K.L.W. wrote the paper. All authors discussed the results and commented on the manuscript. The study was performed under the supervision of P.K.A. and K.L.W.. G.Q.Y. and P.U. contributed equally to this research.

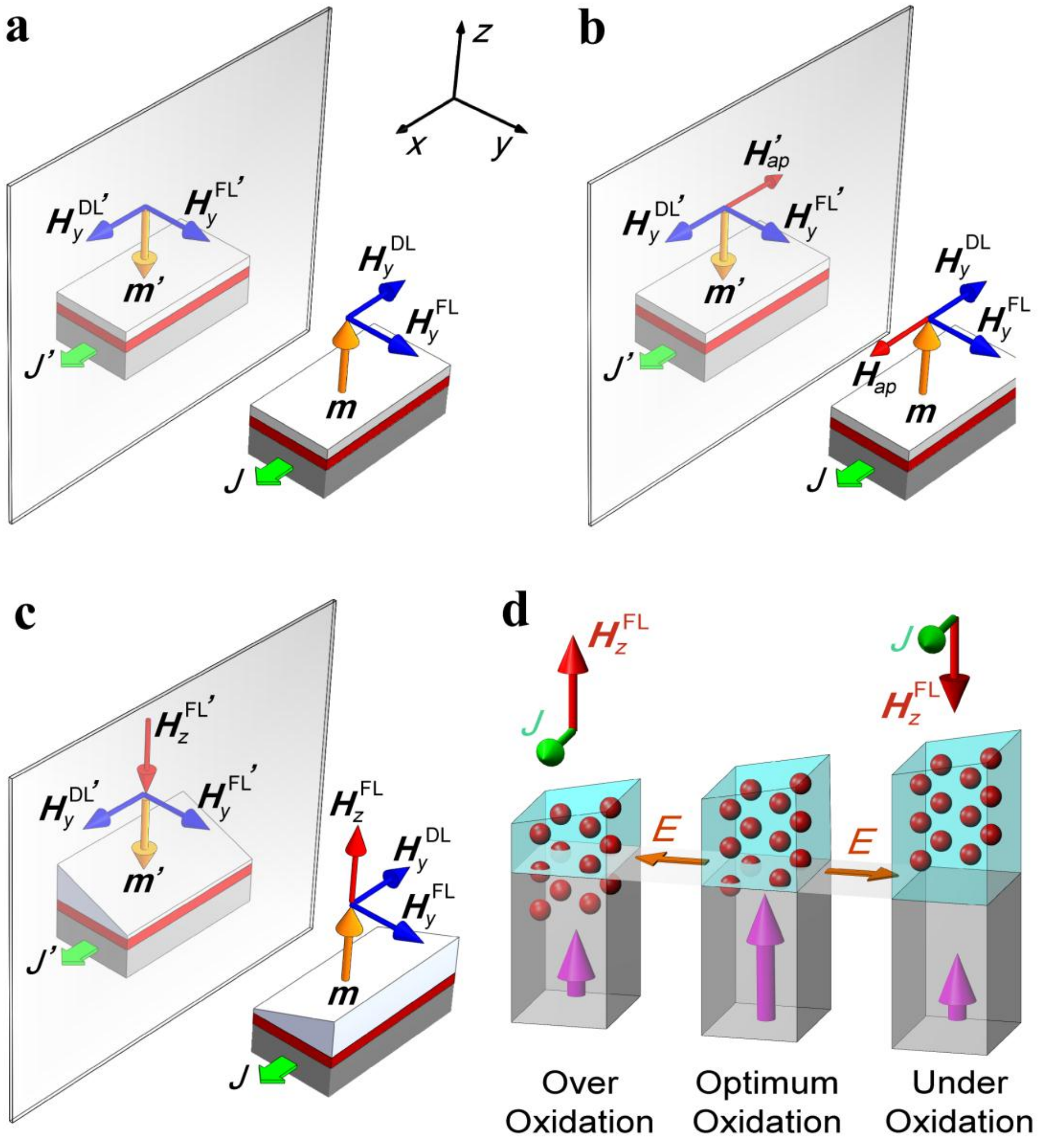

**Figure 1 Schematics of mirror symmetry and current-induced effective fields corresponding to spin-orbit torques (SOTs). a,** Effective fields induced by current, and absence of deterministic out-of-plane switching, in a perpendicular magnetic structure which is symmetric about the *xz* plane. The same direction of current allows for states with magnetization pointing up as well as down. The in-plane effective fields ($\boldsymbol{H}_y^{\mathrm{FL}}$ and $\boldsymbol{H}_y^{\mathrm{DL}}$) are permitted due to the structural inversion asymmetry along the *z*-axis. The blue arrows indicate the $\boldsymbol{H}_y^{\mathrm{FL}}$ and $\boldsymbol{H}_y^{\mathrm{DL}}$ and their mirror reflections $\boldsymbol{H}_y^{\mathrm{FL}\prime}$ and $\boldsymbol{H}_y^{\mathrm{DL}\prime}$ with respect to the *xz* plane. It is important to note the difference between an external field $\boldsymbol{H}_{\boldsymbol{ap}}$ and the damping-like field $\boldsymbol{H}_y^{\mathrm{DL}}$; the latter does not

break the mirror symmetry about the *xz* plane as it depends on $\boldsymbol{m}$ and changes sign when magnetization is reversed, while the former does break the aforementioned symmetry (see **b**). **b**, Symmetry with respect to the *xz* plane is broken by applying an external magnetic field ($\boldsymbol{H_{ap}}$) along the *x*-direction. By fixing the direction of $\boldsymbol{H_{ap}}$ (either positive or negative with respect to the *x*-axis), a unique perpendicular magnetization state can be chosen for each current direction. **c**, The new perpendicular effective field ($\boldsymbol{H}_z^{\mathrm{FL}}$) induced by the laterally asymmetric structure, and its mirror image ($\boldsymbol{H}_z^{\mathrm{FL}\prime}$). The presence of the new perpendicular effective field, induced by the lateral symmetry-breaking, uniquely determines the *z*-component of the magnetization for a particular direction of current, thereby allowing deterministic switching without external magnetic fields. **d**, Schematic representation of the ferromagnet-oxide interface, illustrating its nonuniform oxygen content. The resulting nonuniform charge distribution may produce in-plane electric fields ($E$) along the interface, which in turn can contribute to the observed $H_z^{\mathrm{FL}}$. The red spheres indicate the oxygen atoms and the perpendicular pink arrows correspond to the perpendicular magnetic anisotropy in the magnetic layer.

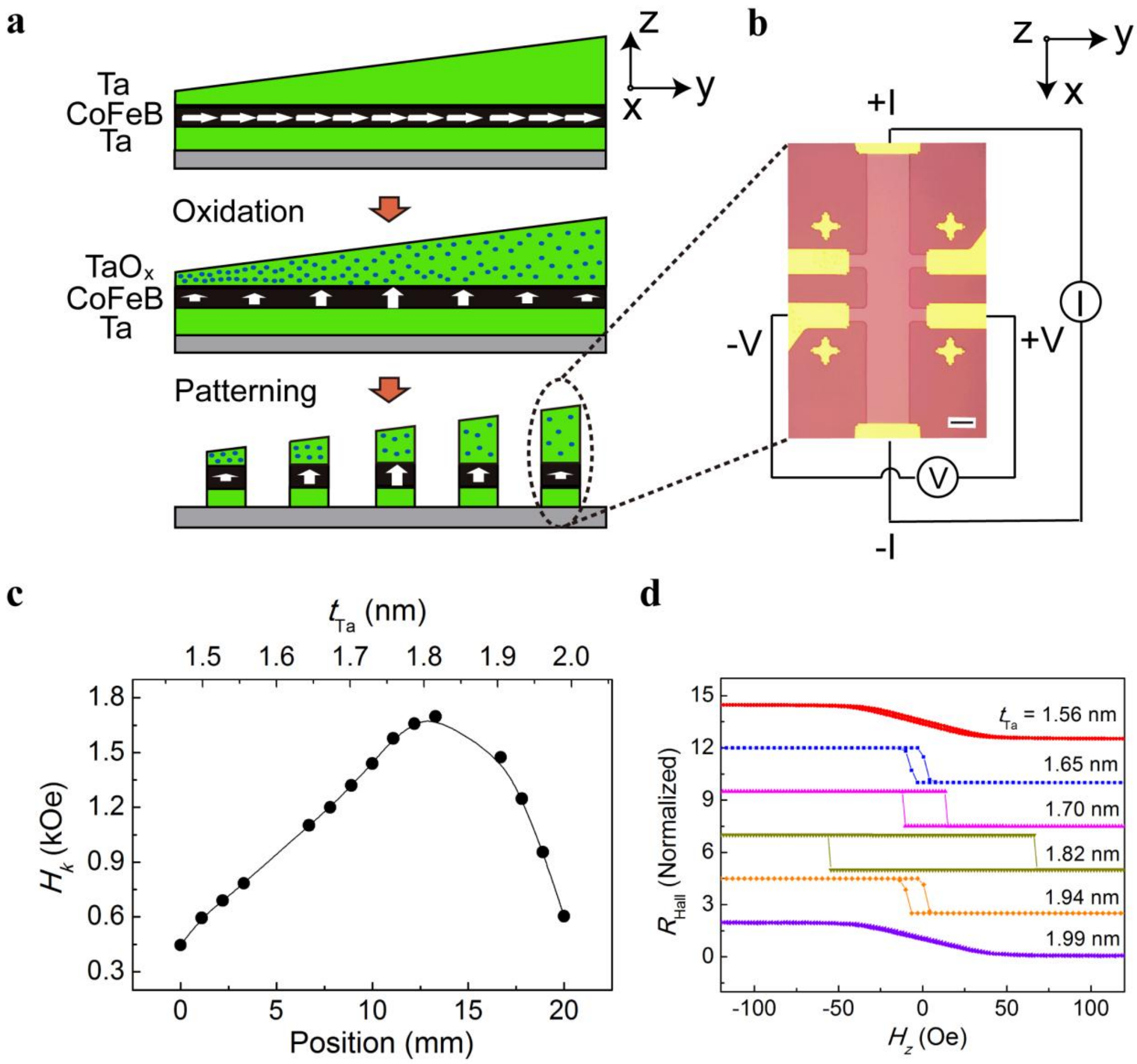

**Figure 2 Device geometry and magnetic perpendicular anisotropy. a**, Procedure for growth and patterning of the devices. The Ta layer on top of the CoFeB film was deposited with a varying thickness across the wafer, resulting in a wedge shape. After the $O_2$/Ar plasma oxidation and annealing, nonuniform perpendicular magnetic anisotropy was realized. **b,** Structure of one device in the array (10 $\mu$m scale bar) and the measurement configuration. Each individual device is designed to have a lateral asymmetry due to the wedge in the $TaO_x$. **c**, Effective perpendicular anisotropy field ($H_k$) as a function of position/thickness of the devices at room temperature. Due to the nonuniform oxidation at the interface, which depends on the thickness of the initially

deposited Ta, a non-monotonic distribution of $H_k$ is obtained. Note that the variation of $H_k$ with respect to position ($dH_k/dy$) can be positive or negative, depending on the device location along the wedge. **d,** Perpendicular magnetization of Ta/CoFeB/$TaO_x$ measured by extraordinary Hall effect (EHE) at different locations along the wedge. A good correlation is found between the out-of-plane coercivity and $H_k$, obtaining a maximum coercivity near the peak of the $H_k$ distribution.

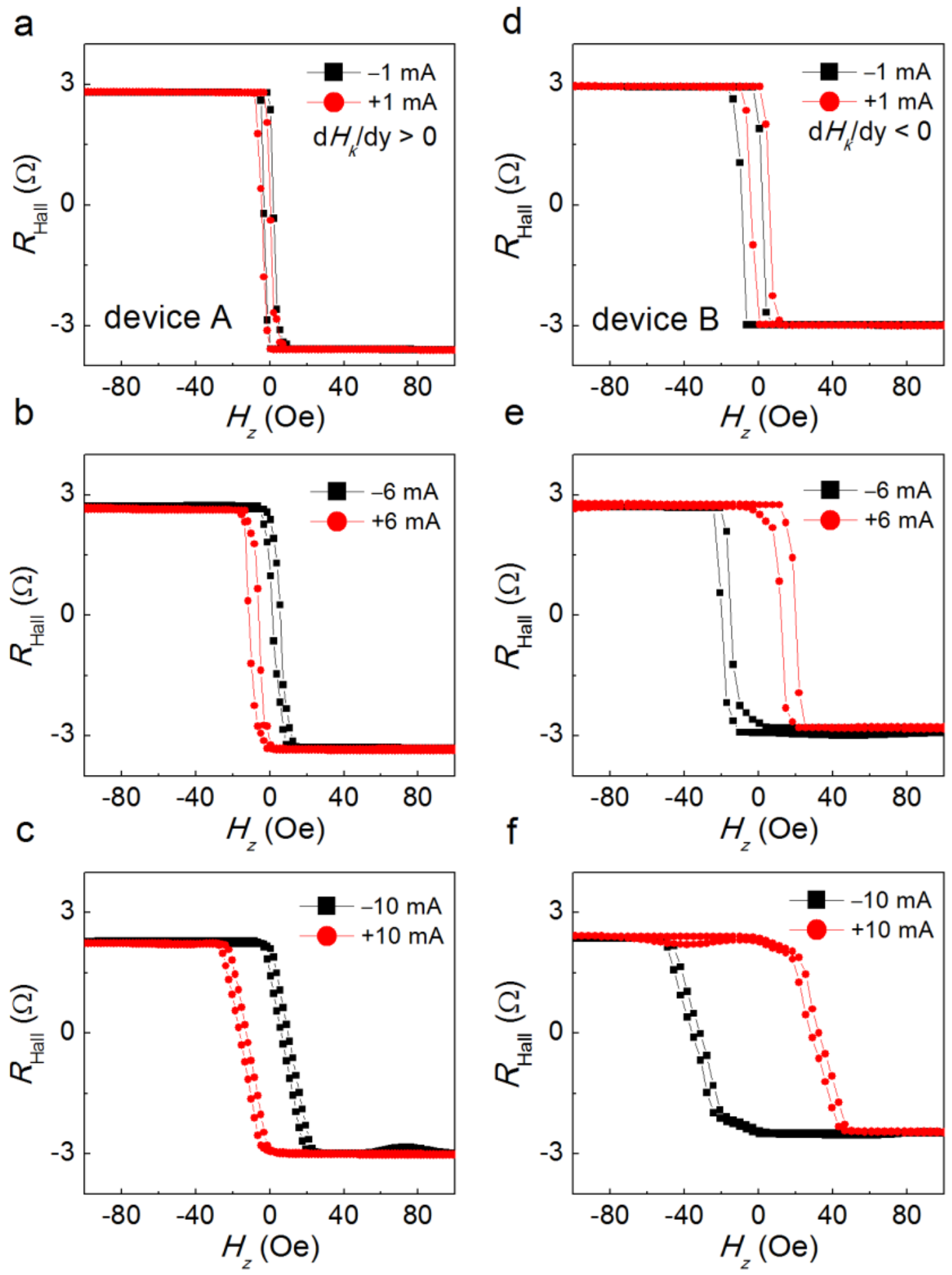

**Figure 3 Effect of $H_z^{\mathrm{FL}}$ induced by current at room temperature.** The figures show the perpendicular magnetization of Ta/CoFeB/$TaO_x$ measured by EHE, while a current of ±1 mA ((**a**) and (**d**)), ±6 mA ((**b**) and (**e**)), and ±10 mA ((**c**) and (**f**)) is applied to the devices. Figures **a-c** show results for a typical device ($t_{\mathrm{Ta}}$ = 1.65 nm before oxidation, device A) in the d$H_k$/d$y$ > 0 region, while **d-f** show results for a typical device ($t_{\mathrm{Ta}}$ = 1.94 nm before oxidation, device B) in the d$H_k$/d$y$ < 0 region. The shift directions of the EHE loops with respect to current, which reflect the directions of $H_z^{\mathrm{FL}}$ induced by the current, are opposite for these two devices with opposite signs of d$H_k$/d$y$.

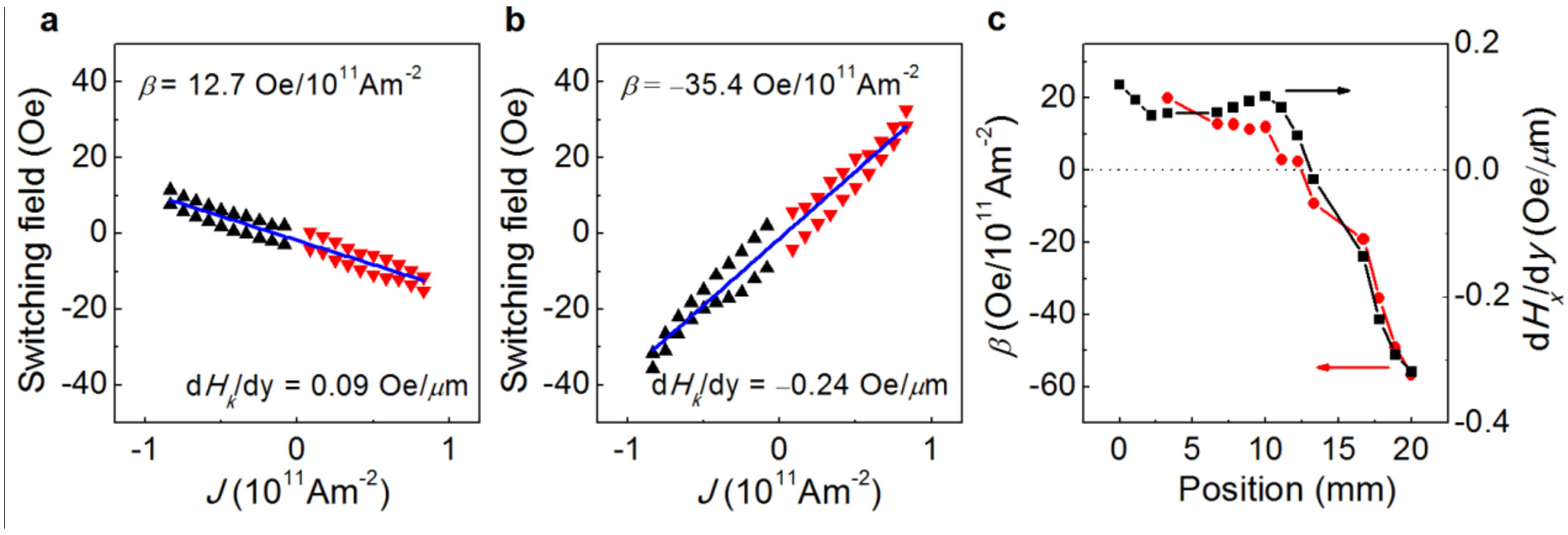

**Figure 4 The switching fields as a function of applied current densities, and $\beta$ as a function of position of the Hall bar devices on the wafer. a, b**, Measured switching fields as a function of current for the devices A (**a**) and B (**b**) discussed in Figure 3, with d$H_k$/dy = 0.09 Oe/$\mu$m and −0.24 Oe/$\mu$m, respectively. The blue lines are linear fits. The values of $\beta$, representing the perpendicular effective field ($H_z^{\mathrm{FL}} = \beta J$), are extracted from the slope. **c,** The red circles correspond to $\beta$ and the black squares correspond to d$H_k$/d$y$, indicating a clear correlation between d$H_k$/d$y$ and $H_z^{\mathrm{FL}}$. This is in agreement with the expected relationship between lateral symmetry breaking and the strength of $H_z^{\mathrm{FL}}$. The dashed line corresponds to zero for both sides of the vertical coordinates.

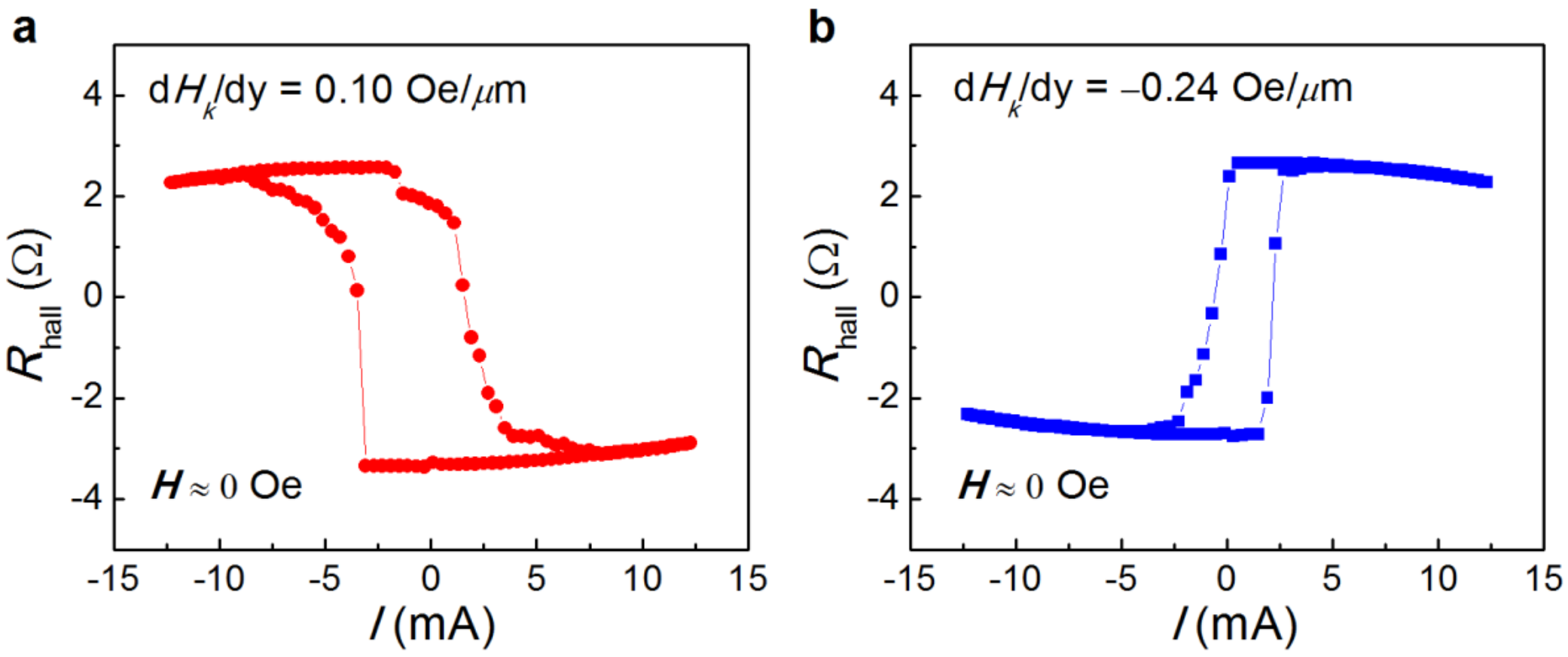

**Figure 5 Switching of perpendicular magnetization by current in the absence of external fields.** Perpendicular magnetization as a function of direct current for devices C ($t_{\mathrm{Ta}}$ = 1.67 nm before oxidation) (**a**) and B ($t_{\mathrm{Ta}}$ = 1.94 nm before oxidation) (**b**) with d$H_k$/d$y$ = 0.10 and −0.24 Oe/$\mu$m, respectively. The favored magnetization direction for each current direction is opposite for these two devices, due to the opposite orientations of the current-induced $H_z^{\mathrm{FL}}$ (i.e. different signs of $\beta$). All measurements were carried out at room temperature.

SUPPLEMENTARY MATERIALS

# Switching of perpendicular magnetization by spin-orbit torques in the absence of external magnetic fields

Guoqiang Yu[1]†, Pramey Upadhyaya[1]†, Yabin Fan[1], Juan G Alzate[1], Wanjun Jiang[1], Kin L. Wong[1], So Takei[2], Scott A Bender[2], Murong Lang[1], Jianshi Tang[1], Yaroslav Tserkovnyak[2], Pedram Khalili Amiri[1] and Kang L. Wang[1]

[1]Department of Electrical Engineering, University of California, Los Angeles, California 90095, United States

[2]Department of Physics and Astronomy, University of California, Los Angeles, California 90095, United States

Table of Contents:

**S1. Symmetry-based model of spin-orbit torques**

In this section, following the symmetry-based phenomenology of deriving the equation of motion for magnetization in the presence of spin-orbit coupling[1], we extend the analysis by Garello *et al.*[2] to derive the form of the new SOT torque terms resulting from additional inversion symmetry breaking along the $y$-axis, i.e Eq. (1) presented in the main text.

In Fig. 1c we schematically show the device geometry having the following symmetry properties. Due to different materials on top and bottom of the ferromagnet, the inversion symmetry along the $z$-axis is broken. Additionally, the wedge of $TaO_x$ on top breaks the inversion symmetry along the $y$-axis, as explained in the main text. In the following we treat the magnetization as a monodomain; however, the analysis can be easily extended to nonuniform magnetization as well[1]. The equation of motion for the magnetization can be written within the Landau-Lifshitz-Gilbert (LLG) phenomenology[3] as

$$\partial_t \boldsymbol{m} = -\gamma \boldsymbol{m} \times \boldsymbol{H} + \alpha \boldsymbol{m} \times \partial_t \boldsymbol{m} + \gamma \boldsymbol{m} \times \boldsymbol{H}^{\mathrm{SOT}}(\boldsymbol{I}, \boldsymbol{m}) \ . \qquad \text{(S1)}$$

Here $\boldsymbol{M} = M_s \boldsymbol{m}$ is the magnetization vector pointing along the unit vector $\boldsymbol{m}$ and having a magnitude $M_s$. The first term on the right hand side represents magnetization precession about an effective field $\boldsymbol{H} = -\delta_{\boldsymbol{M}} F$, derived from a free energy density ($F$) while the second term describes damping of magnetization towards the equilibrium with $\alpha$ and $\gamma$ being the Gilbert damping and gyromagnetic ratio, respectively. The last term corresponds to the current-induced SOT, parameterized by a current- and magnetization-dependent spin orbit field, $\boldsymbol{H}^{\mathrm{SOT}}(\boldsymbol{I}, \boldsymbol{m})$. In the following we derive the form of $\boldsymbol{H}^{\mathrm{SOT}}(\boldsymbol{I},\boldsymbol{m})$ allowed by the symmetries of the structure described above. We take all of the coefficients to be linear in the current *I*. In the presence of structural asymmetries in the *y* and *z* directions, only mirror symmetry in the *x*-direction remains. In order for the equation of motion, Eq. (S1), to remain invariant under this symmetry, the field $\boldsymbol{H}^{\mathrm{SOT}}$ must transform as a pseudo-vector under reflection in the *yz*-plane. We expand the field $\boldsymbol{H}^{\mathrm{SOT}}$ to linear order in $\boldsymbol{m}$. The zero order contribution must be of the form

$$\boldsymbol{H}_0^{\mathrm{SOT}} = H_y^{FL}\boldsymbol{y} + H_z^{FL}\mathbf{z} \ , \tag{S2}$$

which changes sign under $\boldsymbol{I} \to -\boldsymbol{I}$, and therefore transforms as a pseudo-vector under reflection in the *yz* plane. The linear contribution to the field (which transforms as a pseudo-vector) can be written

$$\boldsymbol{H}_1^{\mathrm{SOT}} = H_y^{DL}\boldsymbol{m}\times\boldsymbol{y} + H_z^{DL}\boldsymbol{m}\times\boldsymbol{z} + \boldsymbol{H_d} \ , \tag{S3}$$

where $\boldsymbol{H_d} = A(m_x\boldsymbol{z} + m_z\boldsymbol{x}) + B\big(m_x\boldsymbol{y} + m_y\boldsymbol{x}\big)$ is a dissipative torque which cannot be written as a damping term. So the total current-induced field is given by $\boldsymbol{H}^{\mathrm{SOT}} = \boldsymbol{H}_{\mathbf{0}}^{\mathrm{SOT}} + \boldsymbol{H}_1^{\mathrm{SOT}}$.

**S2. Perpendicular magnetic anisotropy in Ta/$Co_{20}Fe_{60}B_{20}$/$TaO_x$ stack films**

Fig. S1 shows the in-plane and perpendicular M-H curves of the Ta(5)/$Co_{20}Fe_{60}B_{20}$(1)/$TaO_x$(3) (thickness in nm) film. Note that the film normal is along the magnetic easy axis. The saturation magnetization as a function of thickness is shown in Fig. S2. Fig. S3 shows the thickness dependence of $K_{eff}\times t_{CoFeB}$ for Ta(5)/$Co_{20}Fe_{60}B_{20}$(*t*)/$TaO_x$ (thickness in nm) film. It is shown that large PMA can be achieved when $t_{\mathrm{CoFeB}}$ is between 0.9 and 1.2 nm. In our experiments, we use the Ta(5)/$Co_{20}Fe_{60}B_{20}$(1)/$TaO_x$(wedge) (thickness in nm) which has sufficient perpendicular anisotropy to result in an out-of-plane magnetization.

**S3. Calculation of Oersted fields induced by current**

In this section, we estimate the magnitude of Oersted fields expected as a result of an asymmetric flow of current as pointed out in the main text. We find that the magnitude of the resultant average Oersted field parallel to the *z*-axis, $H_{eff}^{O}$, is at least one order of magnitude smaller than the largest effective magnetic field measured in the experiment, hence providing more evidence for a spin-orbit induced mechanism of the observed current-induced field. Given the length of the Hall bar is much larger than its width, which in turn is much larger than the thickness, we calculate the resultant Oersted field due to an infinitely long sheet of width $d$ and zero thickness (see Fig. S4). In this case, the strength of the *z*-component of the Oersted field, $H_z^O$, at a distance $r$ from the center of the sheet due to current $I$ flowing along the *x*-axis is given by

$$H_z^O = \frac{I}{2\pi d} \int_{r-d/2}^{r+d/2} \frac{1}{y} dy = \frac{I}{2\pi d} \log(\frac{r+d/2}{r-d/2}) \ . \qquad \text{(S4)}$$

As a measure of the strength of the net $z$-component of the Oersted field seen by the device we average the above expression over the width where no current flows, (the field for the part where the current flows averages to zero) and taking the extreme case of current flowing through only one half of the width, i.e. $d = w/2$. We obtain the effective Oersted field $H_{eff}^{O} = 0.3$ Oe for the experimental widths of $w = 20\mu m$ and a current of $I$ = 1mA. This value is ~16 times smaller than the maximum field of 4.7 Oe measured experimentally, and hence strongly suggests a microscopic origin different from Oersted fields.

**S4. Second-harmonic measurements for independent characterization of the current-induced perpendicular field-like term $H_z^{\mathrm{FL}}$**

In this section, we apply the small-signal harmonic measurement technique to measure field-like torque due to lateral symmetry breaking. The motivation behind these experiments is to show the presence of the new field-like torque even for the case when the sample is forced to behave as a single-domain particle, ruling out any intricate micro-magnetic switching process as an explanation for the observed shifts in the EHE loops presented in the main text. To this end, different from harmonic measurements found in literature[4, 5], we perform measurements as a function of the strength of the magnetic field whose magnitude is kept greater than the saturation field throughout, ensuring single domain behavior.

In the small signal harmonic measurement scheme, the system is excited by an alternating current $I = I_0 \sin \omega t$, while simultaneously measuring the Hall voltage $V_H \equiv IR_H$ components at first ($V_\omega$) and second harmonic ($V_{2\omega}$). These harmonic components can be obtained by Taylor expanding Hall resistance up to linear order in $I$, i.e. $R_H(I) = R_H^0 + (dR_H/dI)I$, as

$$\begin{gathered} V_\omega(t) = V_\omega \sin\omega t\,, \\ V_\omega = I_0 R_H^0\,, \\ V_{2\omega}(t) = V_{2\omega}\cos 2\omega t\,, \\ V_{2\omega} = -\left(\frac{I_0^2}{2}\right)\left(\frac{\mathrm{d}R_H}{dI}\right). \end{gathered} \tag{S5}$$

The measurement of second harmonic signal thus provides a method to measure $dR_H/$

$dI$, which in turn encodes information about SOTs. The dependence of $dR_H/dI$ on the strength and the orientation of external field has been used to quantify SOTs[4, 5] and their angle dependences[2] for the case when symmetry is broken along the growth direction. In the following, we apply the scheme to measure the additional SOT terms allowed by symmetry for the case of lateral symmetry breaking.

The experiments were performed in a Physical Property Measurement System (PPMS) by rotating the sample. The external magnetic field is rotated in the *yz* plane, making an angle $\theta_B$ with the $z$-axis. An *a.c.* current is applied with constant frequency $\omega$ = 13.931 Hz. Two lock-in amplifiers are used to read the in-phase first harmonic $V_\omega$ and the out of phase second harmonic $V_{2\omega}$voltages. The first harmonic signal (normalized by its maximum) along with the fit to a single domain model (the numerical solution to Eq. (S13)), are shown in Fig. S5. The excellent match between the model and the experimental data proves the single domain behavior in these measurements.

In section S6 we present a general method to derive expressions for the second harmonic signal in the presence of the new SOT terms. These new terms are extracted from $V_{2\omega}$ for the case when the magnetic field, of strength greater than the saturation field, is applied along the *y*-axis. In this case, the dimensionless second harmonic signal, obtained by normalizing $V_{2\omega}$ by $V_\omega^\pi$ (defined as the first harmonic voltage at $\theta_B = \pi$), reads

$$\frac{V_{2\omega}}{V_\omega^\pi} = -\frac{1}{2}\left[\frac{H_z^{FL}}{H_{ap} - H_k} + \frac{R_P(H_z^{DL} + B)}{R_E H_{ap}}\right] \tag{S6}$$

upon substituting $\overline{m}_y = 1$ and $\overline{m}_x = \overline{m}_z = 0$ in Eq. (S15) and using Eq. (S5). Here,

$R_P$ and $R_E$ are the planar and extraordinary Hall coefficients, respectively. It is clear from the above equation that, for this particular configuration, the contributions to second harmonic signal from terms due to inversion symmetry breaking along the $z$-axis drops out, revealing the new terms. This is a consequence of the fact that for equilibrium magnetization pointing along the $y$-axis, torque from terms due to symmetry breaking along the $z$-axis is zero. The dependence (as obtained in Eq. S6 above) on the new fields and the strength of the external field can also be understood within a simple physical picture as the following. The second harmonic signal is proportional to the change in resistance due to deviations from an equilibrium orientation resulting from current-induced torques on the magnetization. This, in turn, has two contributions. First, the new field-like term causes magnetization to deviate in the *yz* plane, contributing a change in the extraordinary Hall resistance (the first term on the right-hand-side of Eq. (S6)). Second, the new dampling-like and dissipative terms cause the magnetization to deviate in the *xy* plane, contributing to a change in resistance via the planar Hall contribution (the second term on the right-hand-side of Eq. (S6)). The deviations from equilibrium are themselves proportional to the strength of the new terms as compared to the effective magnetic field strength keeping the magnetization aligned along the equilibrium direction. The stronger the external magnetic field, smaller the deviations (which ultimately are expected to go to zero for very large strength of external field) hence showing the $1/H_{ap}$ dependence.

In Fig. S6, we present the measured normalized second harmonic signal, as a function of the strength of the external magnetic field at $\theta_B = \pi/2$ and a current with an RMS value of 10mA, for four representative devices along the wedge, each showing $1/H_{ap}$ dependence. In the present second harmonic experiments, the observed signal can originate from both the new field-like and damping-like terms (see Eq. (S6)). However, in order to make quantitative comparison with the "field-like" shifts seen in the EHE loops we make use of the following. For the present material system, since $R_P \ll R_E$ we drop the planar Hall contribution in Eq. (S6) giving[6]

$$\frac{V_{2\omega}}{V_{\omega}^{\pi}} = -\frac{1}{2}\left[\frac{H_z^{FL}}{H_{ap} - H_k}\right]. \tag{S7}$$

The fits of Eq. (S7) to the measured data are also shown in Fig S6, exhibiting an excellent match for these devices. More importantly, the extracted field per unit current density for all of these devices (i.e. $\beta$ = 11.4, ~ 0, −25.8 and −52.2 Oe per $10^{11}$Am$^{-2}$) compares perfectly with the corresponding values ($\beta$ = 12.7, 2.4, −19.1 and −56.7 Oe per $10^{11}$Am$^{-2}$) obtained from the shifts in EHE loops presented in the main text in both the magnitude and the sign. This provides unambiguous evidence of new terms even for single domains.

## S5. Second-harmonic measurements for characterization of the regular current induced field-like term $H_y^{\mathbf{FL}}$

To compare the current-induced perpendicular effective field $H_z^{\mathrm{FL}}$, due to the breaking of inversion symmetry along the $y$-axis, with that of the usual in-plane effective field $H_y^{\mathrm{FL}}$, we look at the slope of second harmonic signal (as measured in section S4) at $\theta_B = \pi$. Again neglecting the planar Hall contribution, differentiating Eq. (S15) with respect to $\theta_B$ and using Eq. (S5), the expression for the slope becomes:

$dV^{2\omega}/d\theta_B = -(I_0 R_E/2)\{H_y^{FL}/(H_{ap}+H_k)\}(d\bar{m}_y/d\theta_B)$. Here we have made use of the fact that at $\theta_B = \pi$, strength of the effective field, as defined in section S6, is $\bar{H}_z' = H_{ap} + H_k$. Solving Eq. (S13) for $\bar{m}_y$ near $\theta_B = \pi$ we get: $d\bar{m}_y/d\theta_B = -H_{ap}/(H_{ap}+H_k)$. Additionally noting $d^2V_\omega/d^2\theta_B = I_0 R_E(d^2\bar{m}_z/d^2\theta_B)$ and $\bar{m}_z \approx -1 + \bar{m}_y^2/2$ (near $\theta_B = \pi$), from the above expression for slope of second harmonic signal we immediately obtain

$$H_y^{FL} = 2H_{ap}\frac{(dV_{2\omega}/d\theta_B)}{d^2V_\omega/d^2\theta_B} \quad . \tag{S8}$$

We note that the new terms do not contribute to the slope at $\theta_B = \pi$, allowing for the extraction of the current-induced field-like term $H_y^{\mathrm{FL}}$. This method and Eq. (S8) is similar to that used in the literature[2, 4, 5], with the only difference being here, rather than varying the strength of external magnetic field, we vary its orientation. In Fig. S7 and S8, we show the first and second harmonic signal near $\theta_B = \pi$ for the device having largest value of the new perpendicular field-like term. The corresponding value of $H_y^{\mathrm{FL}}$ extracted from Eq. (S8) reads 170 Oe per $10^{11}$ $\mathrm{Am}^{-2}$.

## S6. General method of deriving $dR_H/dI$ for the second harmonic measurements

We derive the expressions for $dR_H/dI$ based on the standard scheme of linearization of LLG supplemented with SOTs, including the new terms allowed by lateral symmetry breaking (Eq. (S2) and (S3)). We note that such a scheme can easily be adopted for any other general form of SOT and effective field. In Fig. 2b, we show the experimental geometry under consideration. In this geometry, the Hall resistance can be written as: $R_H = R_E m_z + R_P m_x m_y$, where $R_E$ and $R_P$ stand for the extraordinary and planar Hall coefficient, respectively, resulting in the following

expression

$$\frac{dR_H}{dI} = R_E \frac{dm_z}{dI} + R_P m_y \frac{dm_x}{dI} + R_P m_x \frac{dm_y}{dI} \ . \tag{S9}$$

Thus for the harmonic measurements we will be interested in finding equilibrium orientation of magnetization as a function of current. This equilibrium orientation, obtained by substituting $\partial_t \boldsymbol{m} = 0$ in the equation of motion for magnetization (S1), is given by

$$\boldsymbol{m} \times \boldsymbol{H} = H_y^{\mathrm{FL}} \boldsymbol{m} \times \boldsymbol{y} + H_y^{\mathrm{DL}} \boldsymbol{m} \times (\boldsymbol{m} \times \boldsymbol{y}) + H_z^{\mathrm{FL}} \boldsymbol{m} \times \boldsymbol{z} + H_z^{\mathrm{DL}} \boldsymbol{m} \times (\boldsymbol{m} \times \boldsymbol{z}) \ . \tag{S10}$$

In the following it will be useful to reorient the coordinate system such that the equilibrium orientation in absence of current $\boldsymbol{m_0} = (\overline{m}_x, \overline{m}_y, \overline{m}_z)$, is aligned with the $z$-axis, defining primed coordinates and rotation matrix $U$, as: $\boldsymbol{m}' \equiv U\boldsymbol{m}; \boldsymbol{m}'_{\boldsymbol{0}} = \boldsymbol{z}$. Rotating Eq. (S10) by $U$ gives

$$\boldsymbol{m}' \times \boldsymbol{H}' = H_y^{\mathrm{FL}} \boldsymbol{m}' \times \boldsymbol{y}' + H_y^{\mathrm{DL}} \boldsymbol{m}' \times (\boldsymbol{m}' \times \boldsymbol{y}')$$

$$+ H_z^{\mathrm{FL}} \boldsymbol{m}' \times \boldsymbol{z}' + H_z^{\mathrm{DL}} \boldsymbol{m}' \times (\boldsymbol{m}' \times \boldsymbol{z}'). \tag{S11}$$

Moreover, for the harmonic measurements presented here, we are interested in the case where the current induced spin orbit fields are much smaller than the effective magnetic field. To this end, we linearize Eq. (S11) around $H_y^{\mathrm{FL}} = H_y^{\mathrm{DL}} = H_z^{\mathrm{FL}} = H_z^{\mathrm{DL}} = 0$ by making following substitutions: $\boldsymbol{m}' = \boldsymbol{m}'_0 + \delta\boldsymbol{m}'(I)$ and $\boldsymbol{H}' = \boldsymbol{H}'_0 + \delta\boldsymbol{H}'(I)$. Here, $\boldsymbol{m}'_0 = \boldsymbol{z}$ and $\boldsymbol{H}'_0 = \overline{H}'_z \boldsymbol{z}$ are the equilibrium magnetization and corresponding effective field in the absence of current, while $\delta\boldsymbol{m}'(I) = \delta m'_x \boldsymbol{x} + \delta m'_y \boldsymbol{y}$ and $\delta\boldsymbol{H}'(I) = (\partial_{\boldsymbol{m}'} H'_x \cdot \delta\boldsymbol{m}')\boldsymbol{x} + (\partial_{\boldsymbol{m}'} H'_y \cdot \delta\boldsymbol{m}')\boldsymbol{y}$ represent the small deviations due to currents, i.e. $|\delta\boldsymbol{m}'|, |\delta\boldsymbol{H}'|/\overline{H}'_z \ll 1$ . Following the standard program of linearization, keeping terms up to first order in $\delta\boldsymbol{m}'$, $H_y^{\mathrm{FL}}$, $H_y^{\mathrm{DL}}$, $H_z^{\mathrm{FL}}$ and $H_z^{\mathrm{DL}}$, we get the following general solution for the deviation

$$\boldsymbol{\delta} m'(I) = C^{-1}D \ , \tag{S12}$$

$$C = \begin{bmatrix} -\dfrac{dH'_y}{dm'_x} & \bar{H}'_z - \dfrac{dH'_y}{dm'_y} \\ \dfrac{dH'_x}{dm'_x} - \bar{H}'_z & \dfrac{dH'_x}{dm'_y} \end{bmatrix},$$

$D =$

$$\begin{bmatrix} \boldsymbol{x}\cdot(\boldsymbol{z}\times\boldsymbol{y}')\{H_y^{\mathrm{FL}} + B\bar{m}_x\} + \boldsymbol{x}\cdot(\boldsymbol{z}\times\boldsymbol{z}')\{H_z^{\mathrm{FL}} + A\bar{m}_x\} + \boldsymbol{x}\cdot(\boldsymbol{z}\times\boldsymbol{x}')\{B\bar{m}_y + A\bar{m}_z\} - H_y^{\mathrm{DL}}\boldsymbol{x}\cdot\boldsymbol{y}' - H_z^{\mathrm{DL}}\boldsymbol{x}\cdot\boldsymbol{z}' \\ \boldsymbol{y}\cdot(\boldsymbol{z}\times\boldsymbol{y}')\{H_y^{\mathrm{FL}} + B\bar{m}_x\} + \boldsymbol{y}\cdot(\boldsymbol{z}\times\boldsymbol{z}')\{H_z^{\mathrm{FL}} + A\bar{m}_x\} + \boldsymbol{y}\cdot(\boldsymbol{z}\times\boldsymbol{x}')\{B\bar{m}_y + A\bar{m}_z\} - H_y^{\mathrm{DL}}\boldsymbol{y}\cdot\boldsymbol{y}' - H_z^{\mathrm{DL}}\boldsymbol{y}\cdot\boldsymbol{z}' \end{bmatrix}.$$

The above solution in conjunction with Eq. (S9) constitute the general expression for obtaining $dR_H/dI$.

As an application of the general method presented above, we develop the expressions for the case of a ferromagnet having a uniaxial anisotropy, with easy axis along $z$-axis, and an external magnetic field applied in the $yz$ plane (as per the experiments presented in this work). The effective magnetic field in this case can be written as: $\boldsymbol{H} = H_{ap}\sin\theta_B\boldsymbol{y} + (H_{ap}\cos\theta_B + H_k m_z)\boldsymbol{z}$ . Here $H_{ap}$ and $H_k$ parameterizes the strength of the applied and the anisotropy field, respectively, while $\theta_B$ represents the angle between the external field and the $z$-axis. The equilibrium orientation, $\boldsymbol{m}_0 = (0, \bar{m}_y, \bar{m}_z)$, and corresponding rotation matrix are then obtained from

$$H_k\bar{m}_y\bar{m}_z = H_{ap}\big(\sin\theta_B\bar{m}_z - \cos\theta_B\bar{m}_y\big); \bar{m}_y^2 + \bar{m}_z^2 = 1; U = \begin{pmatrix} 1 & 0 & 0 \\ 0 & \bar{m}_z & -\bar{m}_y \\ 0 & \bar{m}_y & \bar{m}_z \end{pmatrix}. \tag{S13}$$

Evaluation of $\boldsymbol{x}'$, $\boldsymbol{y}'$, $\boldsymbol{z}'$ and $dH'_i/\mathrm{d}m'_j$ with above $U$ followed by substitution in Eq. (S12) gives

$$\delta m'_x = \left[\left(H_y^{DL} - A\right)\bar{m}_z - (H_z^{DL} + B)\bar{m}_y\right]/\bar{H}'_z,$$
$$\delta m'_y = \left[(H_z^{FL} + A\bar{m}_x)\bar{m}_y - \left(H_y^{FL} + B\bar{m}_x\right)\bar{m}_z\right]/(\bar{H}'_z - H_k\bar{m}_y^2). \tag{S14}$$

We evaluate $dm_i/dI$ from Eq. (S14) after rotating it by $U^{-1}$, back to the unprimed coordinate system. Finally, substitution of $dm_i/dI$ in Eq. (S9) gives the main result of this section

$$I_0\frac{dR_H}{dI} = \left(\frac{R_E\left(H_y^{\mathrm{FL}} + B\bar{m}_x\right)}{\bar{H}'_z - H_k\bar{m}_y^2} + \frac{R_P(H_y^{\mathrm{DL}} - A)}{\bar{H}'_z}\right)\bar{m}_y\bar{m}_z$$
$$- \left(\frac{R_E(H_z^{\mathrm{FL}} + A\bar{m}_x)}{\bar{H}'_z - H_k\bar{m}_y^2} + \frac{R_P(H_z^{\mathrm{DL}} + B)}{\bar{H}'_z}\right)\bar{m}_y^2 \ . \tag{S15}$$

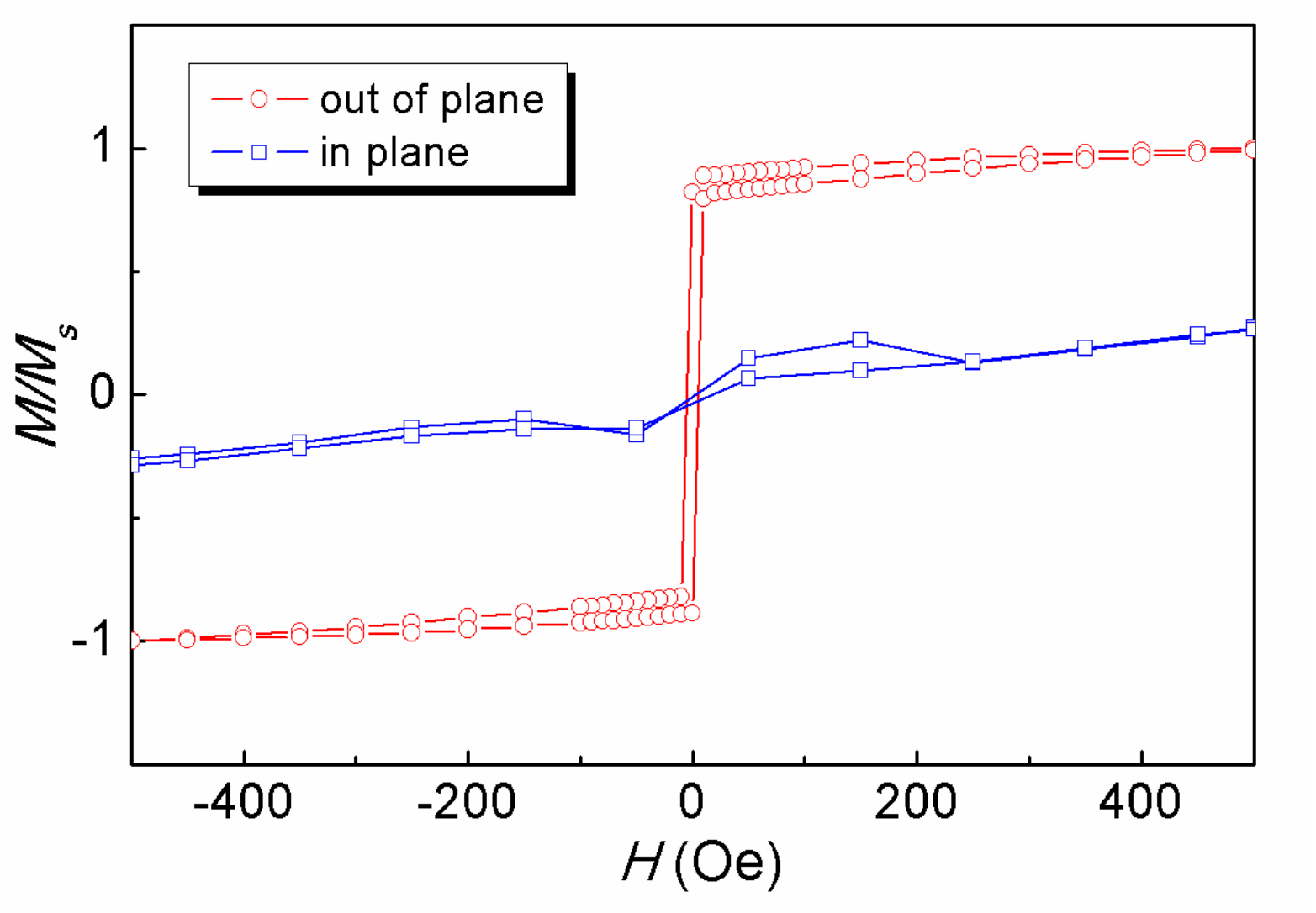

Figure S1. Normalized in-plane and out-of-plane M-H loops for $Ta(5)/Co_{20}Fe_{60}B_{20}(1)/TaO_x$ (thickness in nm) film.

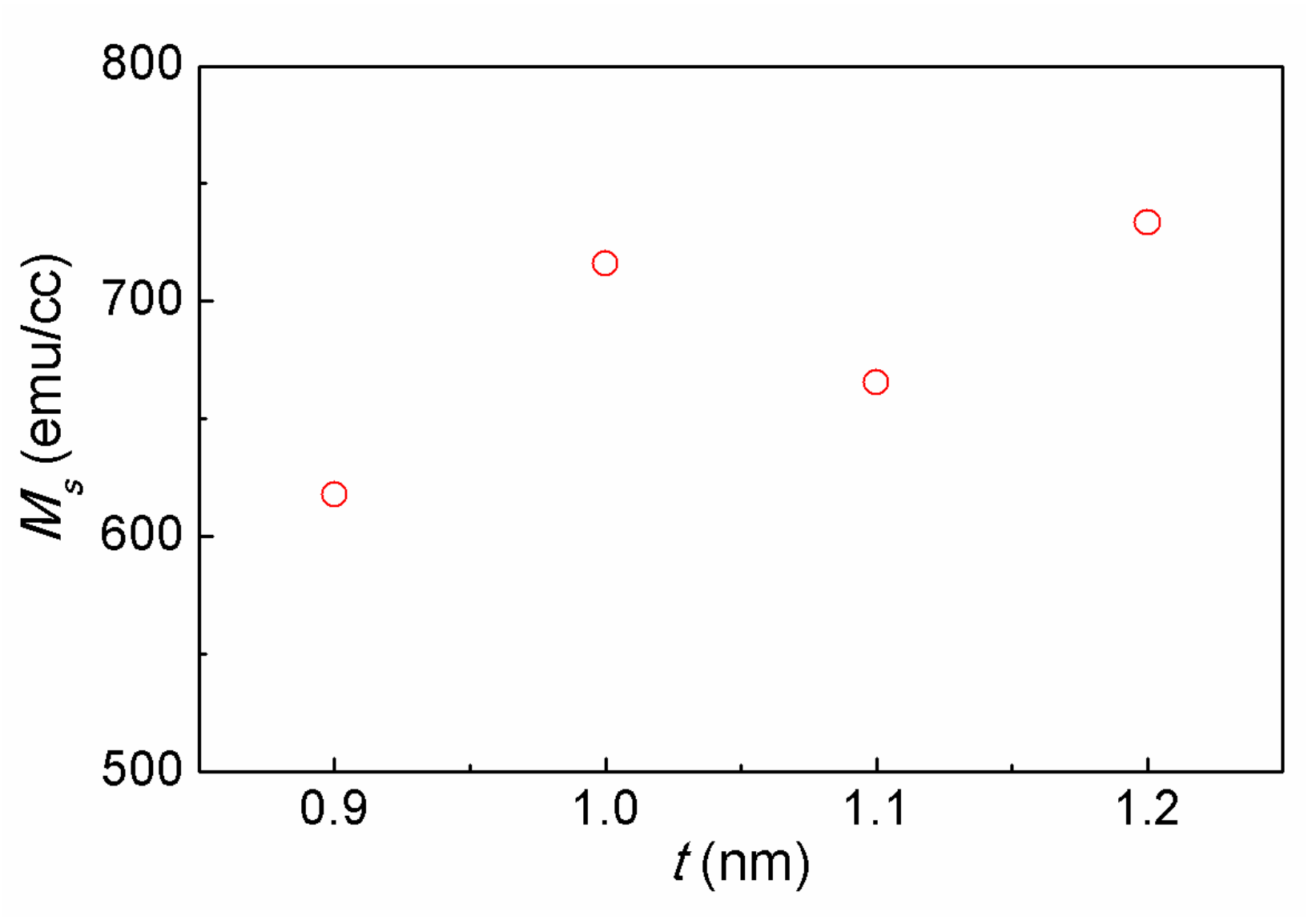

Figure S2. Thickness dependence of $M_s$ for $Ta(5)/Co_{20}Fe_{60}B_{20}$($t$ = 0.9, 1.0, 1.1 and 1.2 nm)/$TaO_x$ (thickness in nm) film.

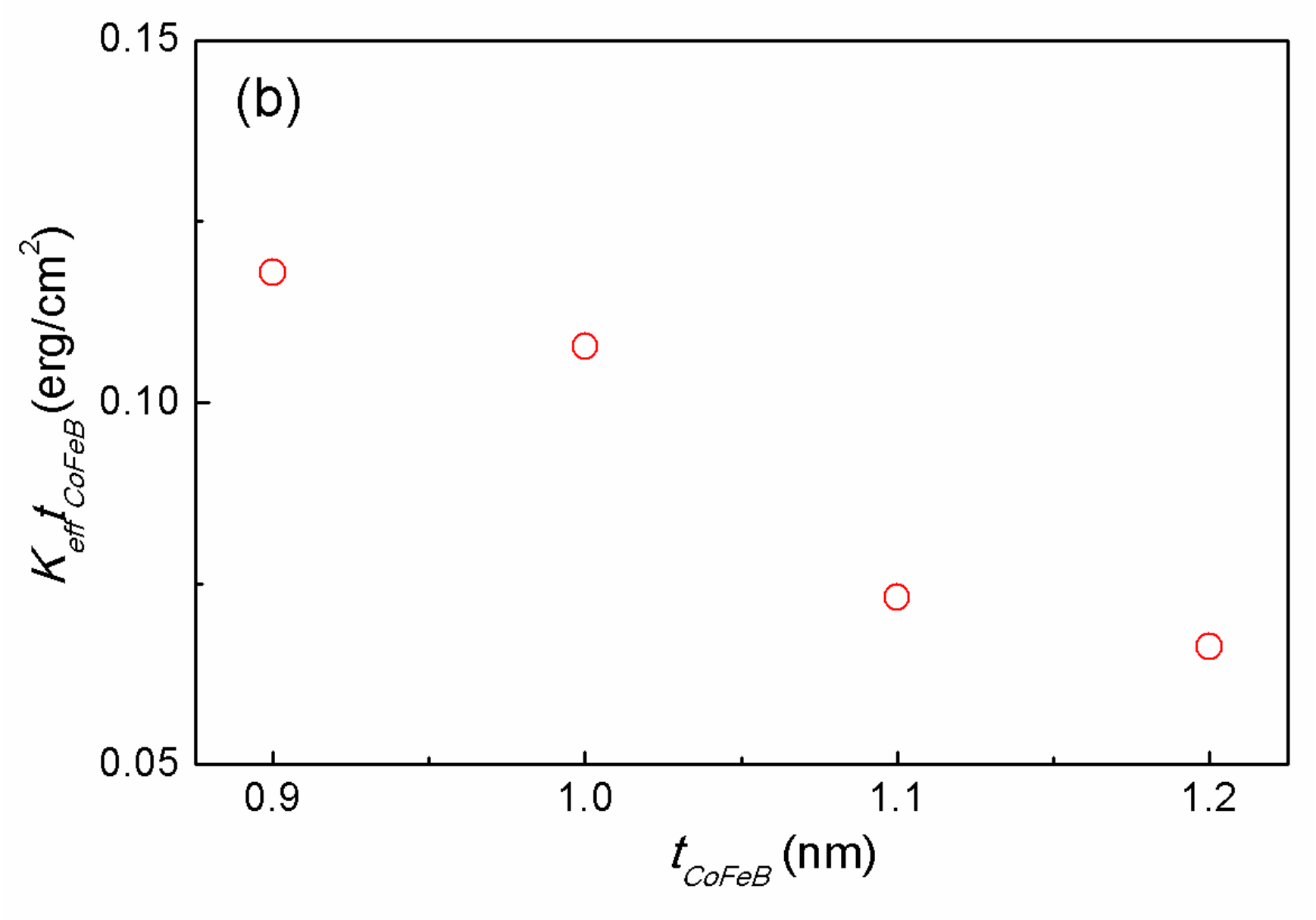

Figure S3. Thickness dependence of $K_{eff}$ $t_{CoFeB}$ for Ta(5)/$Co_{20}Fe_{60}B_{20}$($t$ = 0.9, 1.0, 1.1 and 1.2 nm)/$TaO_x$ (thickness in nm) film.

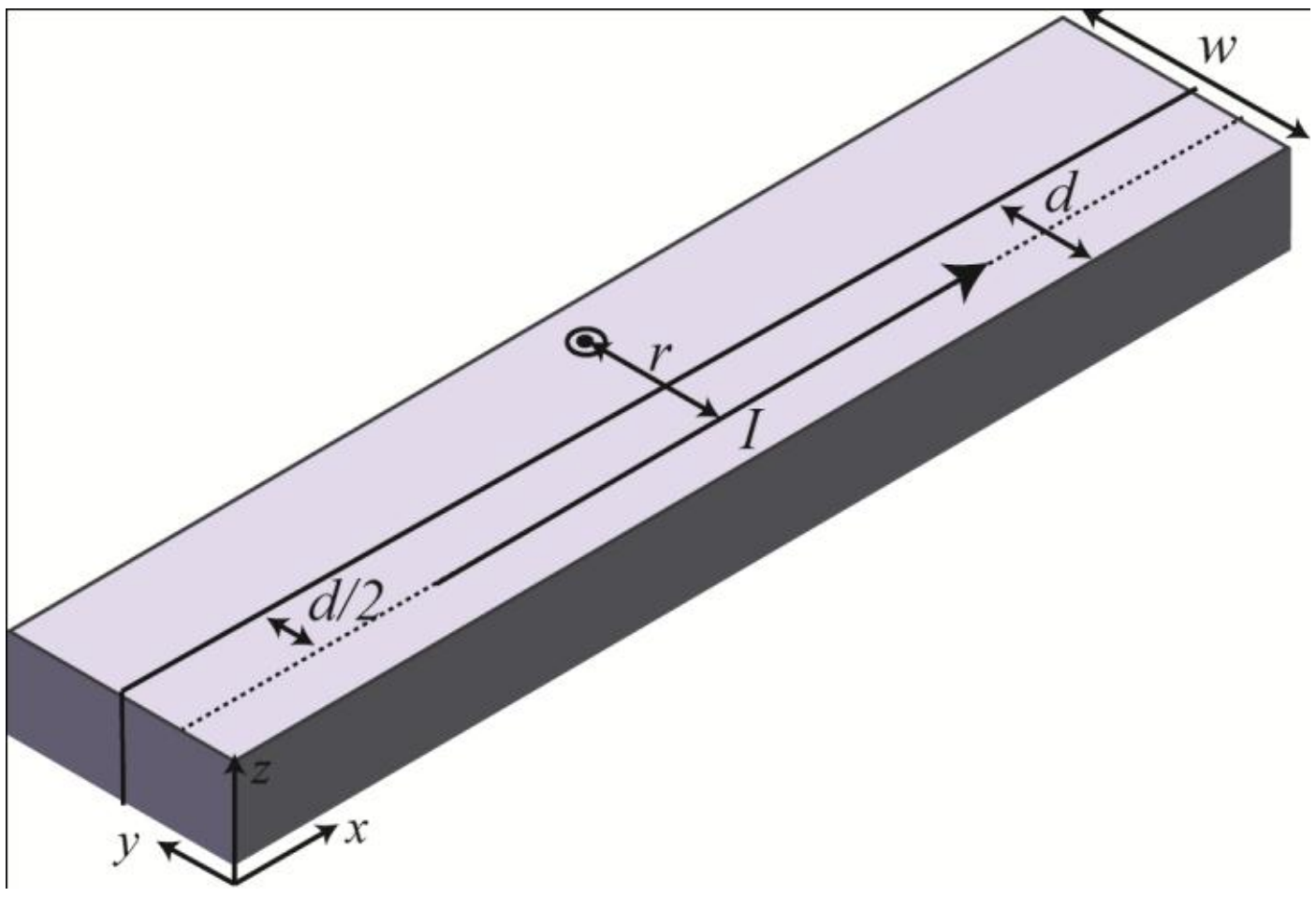

Figure S4. The configuration used to estimate the Oersted field magnitude. Current is assumed to flow only in a region of width $d$. The length of the bar is assumed to be infinite.

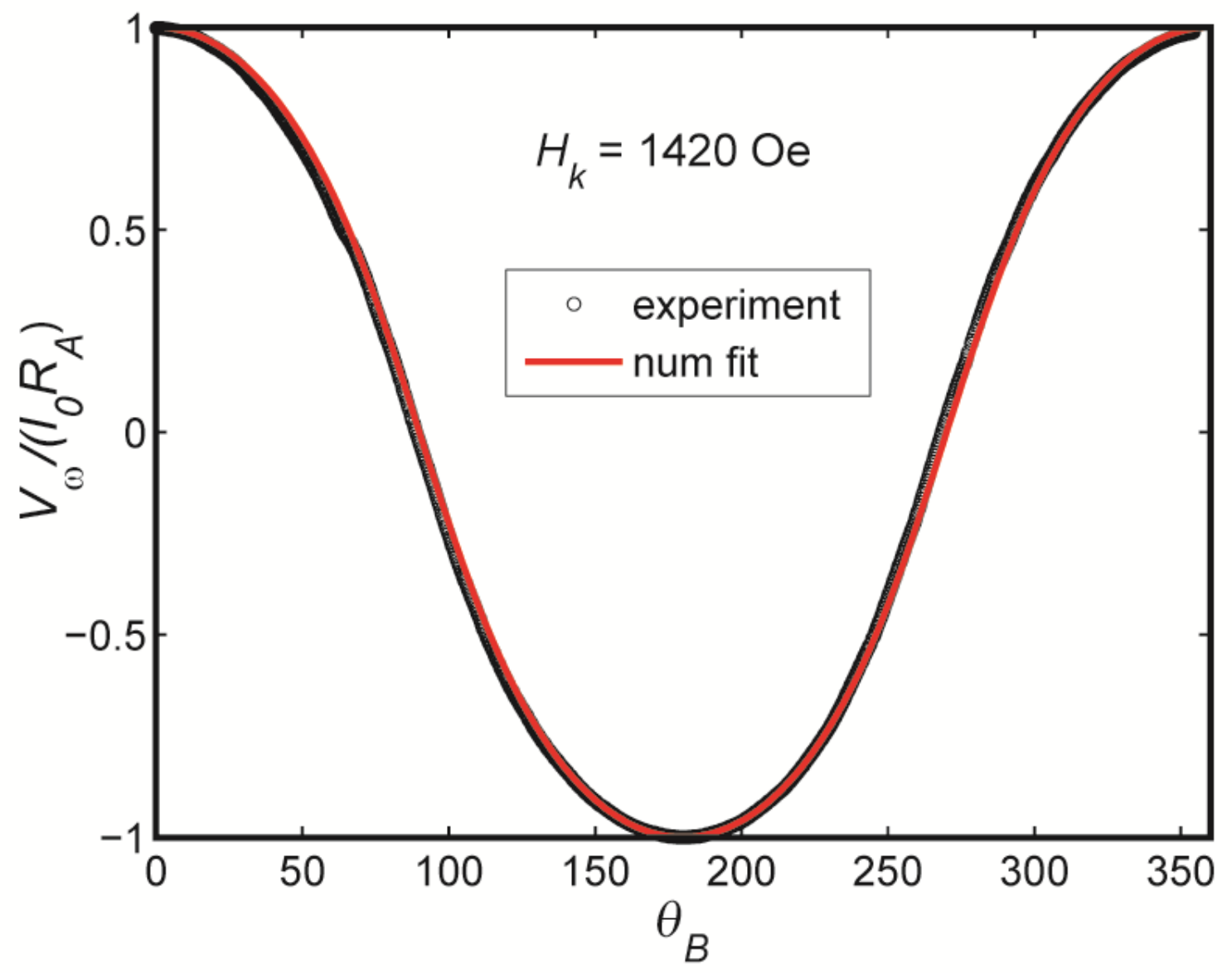

Figure S5. Fit of the single-domain model presented in section S6 to that of the first harmonic signal for a representative device F ($t_{Ta}$ = 1.79 nm before oxidation), extracting the perpendicular anisotropy field $H_k$ .

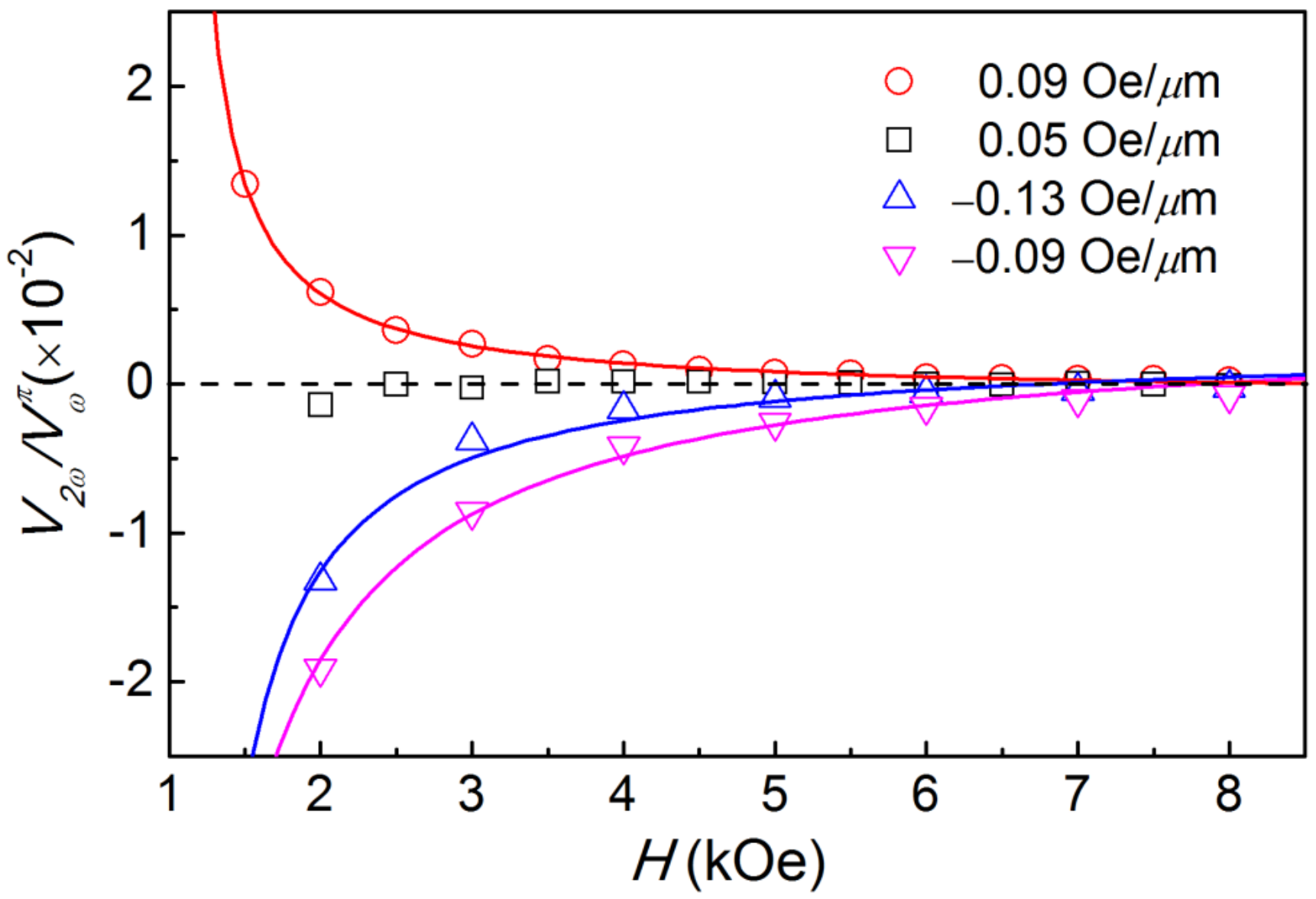

Figure S6. Normalized second-harmonic signal for four representative devices along the wedge. The legend marks the position of the device along the wedge in terms of $dH_k/dy$ as mentioned in the main text.

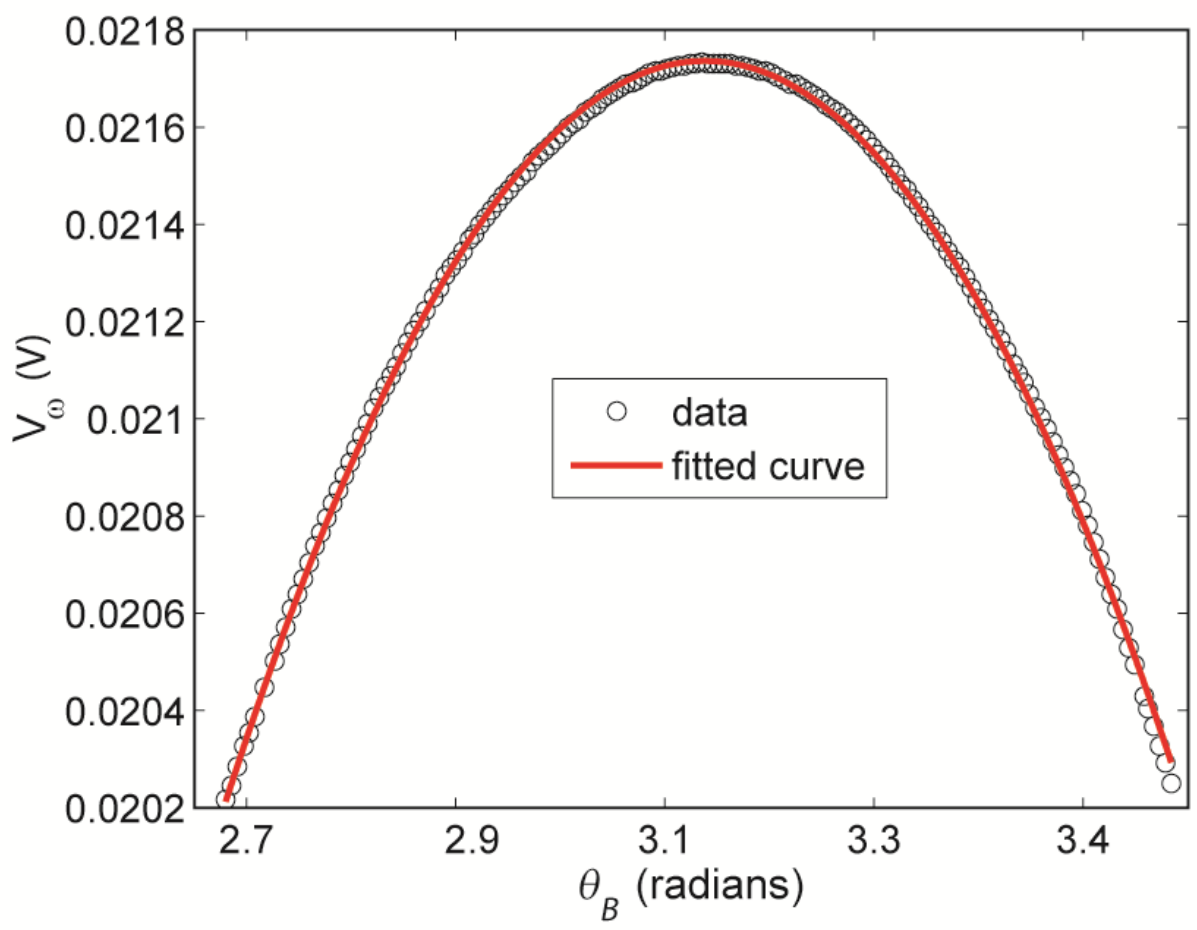

Figure S7. Open symbols represent the first harmonic signal near $\theta_B$ = 180° for the device showing largest value of the current-induced perpendicular field. A parabolic fit to the data is also shown, which is used for extraction of the regular field like term along *y*-axis.

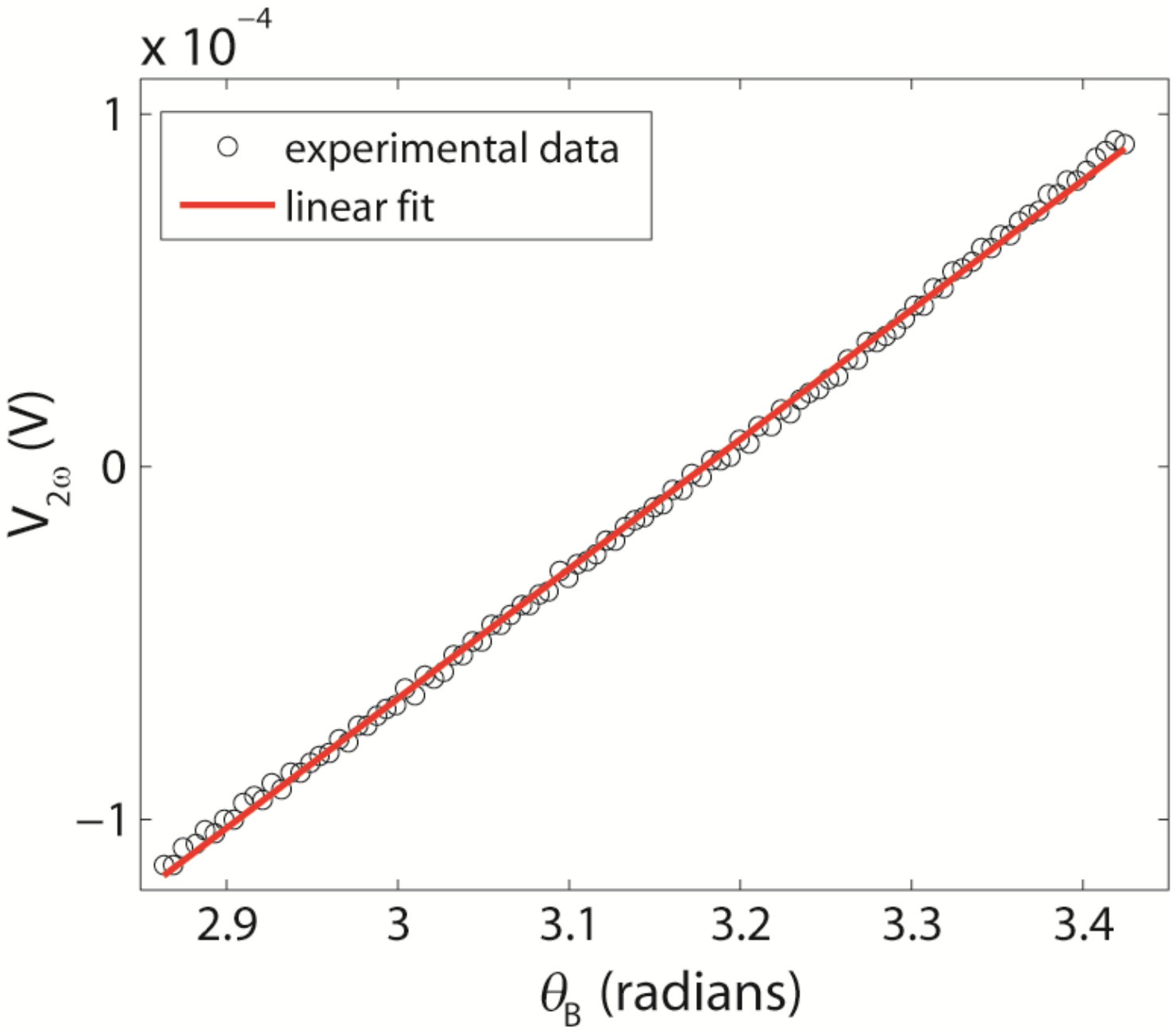

Figure S8. Open symbols represent the second-harmonic signal near $\theta_B$ = 180° for the device showing largest value of the current-induced perpendicular field. A linear fit to the data is also shown, which is used for extraction of the regular field like term along *y*-axis.